\documentclass[aps,prb,showpacs,twocolumn,floats]{revtex4}
\usepackage{amssymb}
\usepackage{amsbsy}
\usepackage{amsmath}
\usepackage{epsfig}

\begin{document}

\title {Spin-Hall effect in triplet chiral superconductors and graphene}

\author{K. Sengupta$^{(1)}$, Rahul Roy$^{(2)}$, and Moitri Maiti$^{(1)}$}

\affiliation{$^{(1)}$Theoretical Condensed Matter Physics Division
and Center for Applied Mathematics and Computational Science \\
Saha Institute of Nuclear Physics, 1/AF Bidhannagar, Kolkata-700064, India.\\
$^{(2)}$University of Illinois, Department of Physics, 1110 W. Green
Street, Urbana, Illinois-61801, USA. }

\date{\today}

\begin{abstract}

We study spin-Hall effects in time-reversal symmetry (TRS) broken
systems such as triplet chiral superconductors and TRS preserved
ones such as graphene. For chiral triplet superconductors, we show
that the edge states carry a quantized spin-Hall current in response
to an applied Zeeman magnetic field $B$ along the ${\bf d}$ vector
\cite{leggett1}, whereas the edge spin-current for ${\bf B} \perp
{\bf d}$ is screened by the condensate. We also derive the bulk
spin-Hall current for chiral triplet superconductors for arbitrary
relative orientation of ${\bf B}$ and ${\bf d}$ and discuss its
relation with the edge spin-current. For TRS invariant system
graphene, we show that the bulk effective action, unlike its TRS
broken counterparts, does not support a SU(2) Hopf term but allows a
crossed Hopf term in the presence of an external electromagnetic
field, which yields a quantized bulk spin-Hall current in response
to an electric field. We also present an analytical solution of the
edge problem for armchair edges of graphene and contrast the
properties of these edge states with their time reversal symmetry
broken counterparts in chiral superconductors. We propose possible
experiments to test our results.

\end{abstract}


\maketitle

\section {Introduction}

The generation and control of spin currents in various solid state
systems have been an area of active research interest. Such systems
can be broadly grouped into two categories; those which have broken
time-reversal symmetry (TRS) and those which do not. Examples of the
first group include chiral superfluids and superconductors such as
two dimensional (2D) $^3$He-A films \cite{leggett1}, 2D layered
strontium ruthenate \cite{maeno1} and $d+id$ superconductor
\cite{senthil1}. In these systems, the superfluid or superconducting
order parameters break TRS and also provide a gap for the
quasiparticle states at Fermi surface. Such a gap, together with
broken TRS, naturally lead to chiral edge states and Hall effect
\cite{ks1,sigrist1,matsumoto1}. However, none of these systems
support the usual charge Hall effect, since they are either charge
neutral ($^3$He-A) or do not have conserved physical charge density
(ruthenates, $d+id$ superconductors)\cite{goryo1,senthil1}. Instead,
they support spin and thermal Hall currents which were found to be
quantized with $N=2$ in $d+id$ systems \cite{senthil1}, and $N=1$ in
$^3$He-A and triplet superconductors \cite{volovik1,stone1}. The
second class of systems which do not break TRS but exhibit spin-Hall
effect include doped semiconductors with spin-orbit interaction
\cite{murakami1} and graphene \cite{kane1}. In graphene, which
provides a realization of Haldane's model \cite{haldane1}, the
spin-orbit interaction provides the gap for the effective low-energy
Dirac-like Hamiltonian of the electrons near the Fermi ($K$ and
$K'$) points \cite{kane1}. Such a gap breaks TRS individually for
each spin species, but leaves the whole system invariant.
Consequently, one expects a quantized charge Hall current $j_{s}$ in
the presence of an external electric field separately for the spin
species $s =\uparrow,\downarrow$. These currents move in opposite
directions so that $j_{\uparrow} = - j_{\downarrow}$. Thus the total
charge Hall current vanishes; however, there exists a net spin-Hall
current $j_{\rm spin} = (j_{\uparrow}-j_{\downarrow})\hbar/2e$. In
the absence of a Rashba term in the Hamiltonian, this spin current
is exactly quantized with a spin-Hall conductivity $\sigma_s =
e/2\pi$. In the presence of the Rashba term, where the number of up
and down spin electrons are not conserved, the spin-Hall
conductivity deviates from its quantized value \cite{haldane2}. This
bulk picture of the spin-quantum Hall effect in graphene is further
supported by numerical verification of the presence of chiral edge
states at the boundaries of the 2D graphene sheet
\cite{haldane2,kane1}. It has also been shown that although the
spin-Hall conductivity deviates from its quantized value, the
spin-Hall effect and the edge states are robust against weak Rashba
coupling \cite{haldane2,kane2}.

The quantization of spin-Hall conductivity in both classes of such
quasi 2D systems necessitates the existence of topological terms in
their low energy bulk effective action. The nature of these
topological terms depends crucially on whether the system is
invariant under TRS or not. For systems with broken TRS, it has been
shown that the low energy effective action has a SU(2) Hopf term
which may lead to quantized spin-Hall current in the presence of an
external magnetic field gradient \cite{volovik1,senthil1,stone1}.
However, the details of the spin response of the triplet
superconductors is expected to be qualitatively different from both
singlet $d+id$ superconductors where the pair potential does not
break spin-rotational symmetry, and $^3$He-A, which, being charge
neutral, do not experience Meissner screening. A detailed
understanding of the spin-current in triplet superconductors
therefore requires an analysis of its effective action in the
presence of an arbitrary Zeeman field. Such a study has not been
undertaken so far. In contrast, for the second class of systems
which respects TRS, one expects a crossed Hopf term \cite{yako1}
which leads to a spin-Hall current in the presence of an external
electric field. However, such a term has not been explicitly derived
for graphene.

The presence/absence of TRS is also crucial for the properties of
the chiral edge states in these systems. Such states are well-known
to exist for both triplet and singlet ($d+id$) chiral
superconductors \cite{ks1,stone1,sigrist1,senthil1}. In these
systems, the edge modes corresponds to localized Bogoliubov
quasiparticles states which have zero charge, but carry a finite
charge current \cite{sigrist1,stone1}. For singlet $d+id$
superconductors, it is well known that the edge states also have a
definite spin quantum number and would therefore carry a quantized
spin-Hall current in the presence of an external magnetic field
\cite{senthil1}. However, the spin structure of the edge states in
triplet chiral superconductors has not been studied before. In
contrast, for systems which preserves TRS, the edge modes, in the
absence of external perturbations, can not carry any net charge
current, but is expected to carry a finite non-quantized spin
current \cite{haldane1,kane1}. However, as we shall show, they can
carry a experimentally detectable quantized charge current, if a
population imbalance of spin-up and spin-down electrons is created
at the edge.

The main results reported in this work are the following. First, for
triplet chiral superconductors with the pair-potential given by
\begin{eqnarray}
\Delta({\bf k}) &=& \Delta_0 \left({\bf s}\cdot {\bf d}\right)
\left(k_x - ik_y\right)/k_F \label{pp1}
\end{eqnarray}
where ${\bf s}$ denotes Pauli matrices in spin-space, $k_F$ denotes
the Fermi wavevector, and ${\bf d}$ denotes a direction in spin
space orthogonal to the spin of the Cooper pair \cite{leggett1}, we
show that the spin structure of the chiral edge states is
qualitatively different from the previously studied $d+id$
superconductors \cite{senthil1}. In particular, these edge states
carry a quantized spin-current in response to an external magnetic
field ${\bf B}$ only if the external magnetic field is applied along
the ${\bf d}$ vector. The spin-current for ${\bf B} \perp {\bf d}$
is screened by Meissner current of the condensate. Second, we obtain
an effective action for the bulk system in the presence of an
arbitrary external Zeeman field, and derive an expression for bulk
spin-current from this effective action for arbitrary relative
orientation of the applied Zeeman field ${\bf B}$ and ${\bf d}$ . We
compare this bulk spin-current with its edge counterpart for both
${\bf B} \parallel {\bf d}$ and ${\bf B} \perp {\bf d}$. In the
limit of zero external Zeeman field our effective action reduces to
those derived in Ref.\ \onlinecite{volovik1, stone1} and contains an
SU(2) Hopf term. Third, for the TRS invariant system graphene,
starting from the low energy Dirac-like Hamiltonian \cite{kane1}, we
derive an effective action in the presence of an external
electromagnetic field, and show that it contains a crossed Hopf term
\cite{yako1} which leads to a quantized spin-Hall current in
response to an external electric field. We also include a weak
Rashba term in the Hamiltonian \cite{kane1,haldane1} and demonstrate
that the spin-current deviates from its quantized value in the
presence of such a term. Finally, we present a solution of the edge
problem in graphene for the armchair edge and obtain analytical
expressions for the energy dispersion and wavefucntions of the edge
states. The properties of the edge states obtained from this
analytical derivation matches previous numerical results
\cite{kane1,haldane1}. We also point out that these edge states can
carry a quantized charge current in response to an applied magnetic
field which can be experimentally measured.

The organization of the paper is as follows. We discuss the
properties of triplet chiral superconductors in Sec.\ \ref{ruth}. In
Sec.\ \ref{ruedge}, we develop the edge state picture and discuss
the spin structure of the edge states. This is followed by Sec.\
\ref{rubulk}, where we obtain a bulk effective action for triplet
chiral superconductors and derive an expression for the spin-Hall
current in these systems. We compare the bulk spin-current with
their edge counterparts for both ${\bf B} \parallel {\bf d}$ and
${\bf B} \perp {\bf d}$. Next, in Sec.\ \ref{graphene}, we study
spin-Hall effect in graphene. In Sec.\ \ref{grbulk}, we derive a
bulk effective action for graphene and show that a crossed Hopf term
exists in its effective action which leads to an quantized spin-Hall
conductivity in the absence of Rashba coupling. We also compute the
deviation of this conductivity from its quantized value in the
presence of a weak Rashba term in the Hamiltonian of the system.
Next, in Sec.\ \ref{gredge}, we present an analytical solution for
the edge state spectrum for the armchair edge and discuss the
properties of these states. This is followed by a discussion of
possible experiments and concluding remarks in Sec.\
\ref{conclusion}.

\section{Chiral triplet superconductors}
\label{ruth}

For triplet superconductors, Cooper-pairing occurs in the $L=1$,
$S=1$ channel, and consequently the pair-potential (Eq.\ \ref{pp1})
breaks spin-rotational invariance. This is manifested in the
presence of the ${\bf d}$ vector in the expression of the
pair-potential (Eq.\ \ref{pp1}) which refers to a direction
orthogonal to the direction of the spin ${\bf S}$ of the Cooper
pair, so that a choice of spin quantization axis along ${\bf d}$
implies opposite-spin pairing \cite{leggett1}. In the presence of
such a pair potential, the action for chiral superconductors can be
written as
\begin{eqnarray}
S &=&  \int d^2r \, dt \psi^{\dagger}(t,{\bf r}) \Big[ i \partial_t
-\tau_3
(\epsilon(-i \nabla)-E_F) \nonumber\\
&&  -\tau_{+} \Delta({\bf k}) -\tau_{-} \Delta^{\ast}({\bf k}) \Big]
\psi(t,{\bf r}),  \label{ac1}
\end{eqnarray}
where ${\bf \tau}$ denotes Pauli matrices acting in particle-hole
space, $\epsilon({\bf k}) = (k_x^2 + k_y^2)/2m$ is the kinetic
energy, $\psi =(\psi_s, g_{ss'} \psi^{\dagger\,s'})$, for spin
indices $s,s'=\uparrow \downarrow$, is the four component Pauli
spinor, and $g_{ss'} = i (s_y)_{ss'}$ is the metric tensor in spin
space \cite{volovik1}. Notice that under time-reversal symmetry
$\Delta(k_x,k_y) \rightarrow \Delta(-k_x,k_y)$ and thus the
pair-potential breaks time reversal symmetry. We shall use this
action to analyze the edge and the bulk properties of the system in
Secs.\ \ref{ruedge} and \ref{rubulk}.

\subsection{Edge states}
\label{ruedge}

In this section, we analyze the spin properties of the edge states.
We shall assume that the ${\bf d}$ vector is fixed along a specific
direction due to the spin-orbit interaction throughout the
superconductors. This is expected to be the case for layered chiral
superconductors ${\rm Sr}_2 {\rm RuO}_4$ where the ${\bf d}$ vector
is fixed along c-axis perpendicular the ${\rm Ru-O}$ plane
\cite{ng1}. In this section, we shall adapt the following convention
for the choice of spin-quantization axis for the electrons: if an
external magnetic field is applied, the spin-quantization will be
chosen along the field; else it will be chosen to be along ${\bf
d}$. For the rest of this work, we shall choose $\hbar=c=1$, unless
explicitly mentioned.

First, we consider the edge problem in absence of the Zeeman field.
Here we shall consider a semi-infinite sample with an edge at $x=0$
and assume a step-function dependence of the pair-potential
\cite{hu1}. In this case, the Bogoliubov-de Gennes (BdG) equation
can be written as
\begin{eqnarray}
  &&\left(\begin{array}{cc}
    {\mathcal H}_0 & \Delta({\bf k_F}) s_z \\
     \Delta^{\ast}({\bf k_F}) s_z & -{\mathcal H}_0
    \end{array}\right) \Psi \left(x,k_{F y}\right) = E \Psi
    \left(x,k_{F y}\right), \nonumber\\
\label{bdg1}
\end{eqnarray}
where ${\mathcal H}_0 = \epsilon(-i \partial_x, k_y) - E_F = -iv_F
\partial_x$ is the linearized dispersion and $k_{Fy}$ is the
in-plane momentum component of the Fermi surface along the edge, and
$\Psi \left(x,k_{F y}\right) = \left( u_s \left(x,k_{Fy}\right), i
(\sigma_y)_{s s'} v^{s'} \left(x,k_{Fy}\right) \right)$ is the
four-component spinor wavefunction. It is well-known that in the
presence of an edge which enforces a boundary condition
$\Psi\left(x=0,k_{Fy}\right)=0$, Eq.\ \ref{bdg1} admits localized
chiral edge states \cite{hu1,ks1,ng1,stone1}
\begin{eqnarray}
\Psi\left(x,k_{F y}\right) &=& 2i \sqrt{\kappa} e^{i k_{F y} y}
e^{-\kappa x} \sin \left(k_{F x} x\right)  {u_{s} \choose {\rm
sgn}(s) v_{{\bar s}}}, \nonumber\\
E\left(k_{Fy}\right) &=& \frac{\Delta_0 k_{Fy}}{k_F}, \quad \kappa =
\frac{\Delta_0}{k_F v_F} k_{Fx}, \label{edgeen1}
\end{eqnarray}
where ${\rm sgn}(s)=\pm$ for $s=\uparrow \downarrow$, $u_s =
exp(i\theta_b) v_{\bar s}$, $\theta_b$ is the global U(1) phase of
the pair-potential $\Delta$, and $u_{\uparrow \downarrow}$ are the
coefficients of spin-up and spin-down components of the normalized
single-particle wavefunctions with $\left|u_s\right|^2 +
\left|v_{\bar s}\right|^2 =1$. These chiral edge quasiparticles are
BCS quasiparticles with equal proportion of spin $s$ electrons and
spin ${\bar s}$ holes; consequently, they do not carry any charge
but have a well defined a spin quantum number ${\rm sgn}(s)/2$ along
${\bf d}$. They have a group velocity $v_{e} =
\partial E\left(k_{Fy}\right)/\partial k_{Fy} = \Delta_0 /k_F$ and
carry a charge current \cite{ng1,stone1,ks2}
\begin{eqnarray}
I_{\rm edge}^{\rm charge} &=& \sum_{k_{Fy},s} \frac{e k_{Fy}}{m}
n_s\left(k_{Fy}\right) = \frac{ek_F^2}{4\pi m}, \label{ic}
\end{eqnarray}
where $n_s\left(k_{Fy}\right)= \theta \left(k_F -k_{Fy}\right)$ is
the density of states for edge state with spin ${\rm sgn}(s)/2$ at
zero temperature. The charge current is not carried with group
velocity and is not quantized. The presence of such a charge current
at the edge of a superconductor also leads to Meissner screening
current by the condensate, so that the sum of magnetic fields due to
both the currents vanish at the bulk of the superconductor
\cite{ks2,matsumoto1,volovik2}.

\begin{figure}
\rotatebox{0}{
\includegraphics[width=\linewidth]{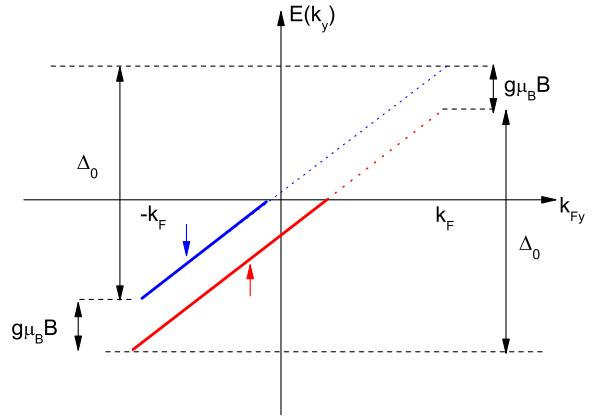}}
\caption{Chiral edge states in spin triplet superconductors in the
presence of a Zeeman magnetic field $B$ along ${\bf d}$. The red
(blue) solid lines denote occupied spin-up (spin-down) edge states
while the corresponding dotted lines show the unoccupied states. The
dashed black lines are guide to the eye. } \label{figspin1}
\end{figure}

In the absence of any external Zeeman magnetic field, the spin-up
and spin-down edge quasiparticle states are equally populated and
hence the spin-current carried by them cancels. Now, let us apply a
Zeeman magnetic field $B$ along ${\bf d}$ so that (choosing the
spin-quantization axis along $B$ and ${\bf d}$) the BdG equations
for the edge quasiparticles states can be written as
\begin{eqnarray}
  &&\left(\begin{array}{cc}
    {\mathcal H}_0 - g\mu_B B s_z & \Delta({\bf k_F}) s_z \\
     \Delta^{\ast}({\bf k_F}) s_z & -\left({\mathcal H}_0 + g \mu_B
     B s_z \right)
    \end{array}\right) \Psi \left(x,k_{F y}\right) \nonumber\\
&&  \quad \quad \quad \quad \quad \quad \quad \quad \quad \quad
\quad \quad \quad \quad \,\,= E \,  \Psi
    \left(x,k_{F y}\right),
\label{bdg2}
\end{eqnarray}
where $\mu_B$ is the Bohr magneton and $g$ denotes the gyromagnetic
ratio. We note that here the effect of the magnetic field is to
shift the energies of the spin-up and spin-down states in {\it
opposite} direction: $E_s \left(k_{Fy}\right)= \Delta_0 k_{Fy}/k_F
-{\rm sgn}(s) g \mu_B B$. As a result, there is an imbalance in the
population of the spin-up and spin-down quasiparticle states as
shown in Fig.\ \ref{figspin1} which leads to a net quantized edge
spin-current
\begin{eqnarray}
(I_{\parallel}^z)_{\rm edge}^{\rm spin}  &=& \frac{1}{2} v_e \left(
N_{\uparrow} - N_{\downarrow} \right) = \frac{1}{8 \pi} g \mu_B B.
\label{is1}
\end{eqnarray}
Note that the spin current, in contrast to the charge current, is
associated with the group velocity $v_e$ of the edge states. Further
with our choice of spin-quantization axis (along ${\bf d}$), the
condensate comprises of Cooper pairs with {\it opposite spin
pairing}. Hence the condensate do not carry any spin or spin-current
and can not screen the edge spin-current.

Such a quantized spin current in the presence of external Zeeman
field is also carried by edge states in singlet $d+id$
superconductors regardless of the direction of the applied magnetic
field. The situation for the triplet superconductors is different as
we now show by studying the spin response of the edge states in the
presence of the Zeeman field applied perpendicular to ${\bf d}$.
Here, without loss of generality,  we choose ${\bf d}={\hat y}$ and
take the spin-quantization axis along ${\bf H}=H {\hat z}$. The BdG
equation can now be written as
\begin{eqnarray}
  &&\left(\begin{array}{cc}
    {\mathcal H}_0 - g\mu_B B s_z & \Delta({\bf k_F}) s_y \\
     \Delta^{\ast}({\bf k_F}) s_y & -\left({\mathcal H}_0 - g \mu_B
     B s_z \right)
    \end{array}\right) \Psi \left(x,k_{F y}\right) \nonumber\\
&&  \quad \quad \quad \quad \quad \quad \quad \quad \quad \quad
\quad \quad \quad \quad \,\, = E \,  \Psi
    \left(x,k_{F y}\right).
\label{bdg3}
\end{eqnarray}
One immediately notes that in this case, the magnetic field does not
shift the energy of the edge states, but shifts the Fermi
wave-vector for the spin-up and spin-down quasiparticles in opposite
directions: $k_{F s} = \sqrt { k_F + {\rm sgn}(s) 2 m g \mu_B B}$.
The BCS quasiparticles are now equal superposition of electron and
hole states with the {\it same spin} $s$. As a result, just as in
the case of charge, they do not carry a spin quantum number along
$z$: $\left<s_z\right> \equiv \left(\left|u_s\right|^2
-\left|v_{s}\right|^2\right) =0$. The edge states now carry a net
charge current
\begin{eqnarray}
(I_{\perp})_{\rm edge}^{\rm charge} &=& \sum_s (I_{s,\perp})^{\rm
charge}_{\rm edge}, \nonumber\\
(I_{s,\perp})^{\rm charge}_{\rm edge} &=& \sum_{k_{Fy}=0}^{k_{Fs}} e
\frac{\hbar k_{Fy}}{m} n_s\left(k_{Fy}\right) =\frac{ek_{Fs}^2}{8\pi
m}. \label{edgecharge}
\end{eqnarray}
Now let us consider Meissner screening of this charge current. Since
we have chosen the direction of spin quantization axis perpendicular
to the ${\bf d}$ vector, we have equal spin pairing in the bulk
($\Delta_{\uparrow \uparrow} = \Delta_{\downarrow \downarrow} =
\Delta$) and hence the Cooper pairs carry a net $s_z = \pm \hbar$.
In the presence of a Zeeman magnetic field, the spin-up and the
spin-down condensates see a shifted chemical potential
$\mu_{\uparrow \downarrow} = \mu + \pm g\mu_B B$. In this situation,
one can apply the arguments of Refs.\ \onlinecite{sigrist1,volovik2}
in a straightforward manner and show that the edge currents would be
screened by spontaneously generated Meissner current
\begin{eqnarray}
I^{\rm Meissner} &=& \sum_s I^{\rm Meissner}_s, \nonumber\\
I_s^{\rm Meissner} &=&  - (I_{s \perp})_{\rm edge}^{\rm charge},
\label{meissner1}
\end{eqnarray}
so as to cancel the net magnetic field due to the edge currents in
the bulk.

The edge states now also carry a net spin current, since the
electron and holes with spin $s$ constituting the BCS quasiparticles
have opposite velocities. In the presence of an external Zeeman
field ${\bf B} \perp {\bf d}$, $k_{F\uparrow}$ and $k_{F\downarrow}$
are different, and consequently the net spin-current is given by
\begin{eqnarray}
(I_{\perp}^z)_{\rm edge}^{\rm spin} &=& \frac{1}{2e} \sum_s {\rm
sgn}(s) (I_{s,\perp})_{\rm edge}^{\rm charge} = \frac{g \mu_B B}{8
\pi}. \label{perpspin}
\end{eqnarray}
Notice that the spin current is not carried by the group velocity.
Further, since the condensate now carry a net spin, the Meissner
current generated as the response to edge charge current also carry
a net spin-current
\begin{eqnarray}
I^{\rm spin}_{\rm Meissner} &=& \frac{1}{2e} \sum_s {\rm sgn}(s)
I_s^{\rm Meissner}, \nonumber\\
&=& - \frac{1}{2e} \sum_s {\rm sgn}(s)(I_{s,\perp})^{\rm
charge}_{\rm edge}, \nonumber\\
&=& - (I_{\perp}^z)_{\rm edge}^{\rm spin}.
\end{eqnarray}
Thus the edge spin current generated by the response of a magnetic
external Zeeman field  ${\bf B} \perp {\bf d}$ is screened by the
condensate, analogous to the charge current. Therefore the response
of the edge states to an applied Zeeman field in triplet chiral
superconductors depends crucially on the direction of the applied
field. We shall discuss possible experiments to probe this behavior
in Sec.\ \ref{conclusion}.

Before ending this section, we note that all of the results derived
can be applied to ruthenates where the pair-potential is expected to
have horizontal line of nodes with the substitution $\Delta
\rightarrow \Delta \left|\cos(ck_z)\right|$ \cite{srrefs}. In
particular, it can be shown that the spin response of the edge
states remain qualitatively similar in the presence of such
horizontal line of nodes.

\subsection{Bulk effective action}
\label{rubulk}

To derive the bulk effective action, we begin with the action of
chiral triplet superconductors in the presence of an external Zeeman
field $B(t,{\bf r})$ and an arbitrary ${\bf d}$ field configuration.
In the presence of such a term the action can be written as
\begin{eqnarray}
S &=&  \int d^2r \, dt \psi^{\dagger}(t,{\bf r}) \Big[ i \partial_t
-C_0^{\rm ext} \nonumber\\
&& -\tau_3
(\epsilon(-i \partial_i -C_i^{\rm ext}) -E_F)
 -\tau_{+} \Delta(-\partial_i -C_i^{\rm ext})\nonumber\\
&&  -\tau_{-} \Delta^{\ast} (-i\partial_i-C_i^{\rm ext}) \Big]
\psi(t,{\bf r}), \label{ac2}
\end{eqnarray}
where $C_0^{\rm ext} = g\mu_B {\bf s}\cdot {\bf B}(t,{\bf r})$ is
the Zeeman field and $\Delta$ is the pair-potential defined in Eq.\
\ref{pp1}. Here, and in the rest of this section, we shall use the
notation $\partial_{\mu} = (\partial_t,
\partial_x,\partial_y)$ and adapt
the sign convention that all covariant vectors have $F_{\mu} =
(F_t,-F_x, -F_y)$ and all operators have $\partial_{\mu} =
(\partial_t,\partial_x,\partial_y)$. The corresponding contravariant
vectors and operators can be obtained by applying the metric tensor
$g_{\mu \nu} = g^{\mu \nu} = (1,-1,-1)$. Also, here we have followed
Ref.\ \onlinecite{volovik1} to treat the Zeeman magnetic field as
the time component of an external SU(2) gauge field $C_{\mu}^{\rm
ext}= 1/2 s_{\alpha} \cdot \Omega_{\mu}^{\alpha\, \rm {ext}}$ with $
{\bf \Omega}_{t}^{\rm ext} = g \mu_B {\bf B}$. The spatial component
of $C_{\mu}^{\rm ext}$ are fictitious fields which shall be set to
zero at the end of the calculation. The advantage of introducing
these fictitious gauge fields becomes apparent when we note that the
spin-current can be obtained using these fields as
\begin{eqnarray}
j_i^{\alpha} &=&  \left< \psi'^{\dagger} \frac{1}{2} \left \{
\partial_{k_{\mu}} G_0^{-1}(k), s^{\alpha}
\right\}_+ \psi'(k) \right>_S,  \nonumber\\
&=&  \frac{\delta S_{\rm eff}}{\delta \Omega_{i}^{\alpha \, {\rm
ext}}}\Big |_{\Omega_{i}^{\alpha \, {\rm ext}}=0}, \label{scu1}
\end{eqnarray}
where $S_{\rm eff}$ is the effective action obtained by integrating
out Fermions in $S$ (Eq.\ \ref{ac2}).

Next we introduce a SU(2) rotation in spin space through the
transformation $\psi(t,{\bf r}) \rightarrow \psi'(t,{\bf r}) =
U^{\dagger}  \psi(t,{\bf r})$, so that the local SU(2) rotation
matrix rotates the ${\bf d}$ vector to z: $U^{\dagger} \left({\bf
s}\cdot {\bf d}\right) U = s_z $. One can now write the action as
\begin{eqnarray}
S &=& S_0 + S_1 \label{ac4}, \\
S_0 &=& \int \frac{d^2k d\omega}{(2\pi)^3} \psi'^{\dagger}(k)
G_0^{-1} \psi'(k), \label{ss0}\\
S_1 &=& -\int \frac{d^2
k d^2p d\omega d p_0 }{(2\pi)^6} \psi'^{\dagger}(k+p) \nonumber\\
&& \times \frac{1}{2} \left \{
\partial_{k_{\mu}} G_0^{-1}(k), C_{\mu}^{\rm total}(p)
\right\}_+ \psi'(k), \label{s1}
\end{eqnarray}
where the notation $\left\{ .. \right\}_{+}$ denotes
anticommutation, the Green function $G_0(k)$ is given by
\begin{eqnarray}
G_0(k;\Delta_0) &=& \frac{\omega + \tau_3 \left(\epsilon({\bf
k})-E_F\right)+ \Delta_0 \left({\bf k}\cdot {\bf \tau}\right) s_z
}{\omega^2 - E_k^2 +i \eta},
\label{green2} \\
E_k &=& \sqrt {\left(\epsilon({\bf k})-E_F \right)^2 + \Delta_0^2},
\label{en2}
\end{eqnarray}
and the gauge fields $C_{\mu}^{\rm total}$ are given by
\begin{eqnarray}
C_{\mu}^{\rm total} &=& U^{\dagger} C_{\mu}^{\rm ext} U +
C_{\mu}^{\rm int}, \nonumber\\
C_{\mu}^{\rm int} &=& -i U^{\dagger} \partial_{\mu} U = \frac{1}{2}
s_{\alpha} \Omega_{\mu}^{\alpha\,{\rm int}}. \label{gf}
\end{eqnarray}
Note that here the internal gauge fields $C_{\mu}^{\rm int}$ are
'pure' gauge fields in the sense that they satisfy
\begin{eqnarray}
\partial_{\mu}{\bf \Omega}_{\nu}^{\rm int} - \partial_{\nu}{\bf \Omega}_{\mu}^{\rm int}
- {\bf \Omega}_{\mu}^{\rm int} \times {\bf \Omega}_{\nu}^{\rm int}
&=& f_{\mu \nu}^{\rm int}= 0. \label{fint1}
\end{eqnarray}
These fields are small when the ${\bf d}$ fields are slowly varying.
Thus for a slowly-varying configuration of both ${\bf d}$ field and
external Zeeman field ${\bf B}$ , one can carry out a gradient
expansion in $C_{\mu}^{\rm total}$.
\begin{figure}
\rotatebox{0}{
\includegraphics[width=\linewidth]{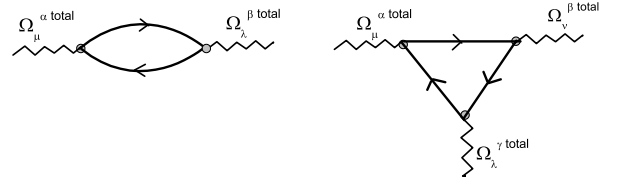}}
\caption{Feynman diagrams which contribute to the SU(2) Hopf terms
effective action. The closed circles represent $ 1/2 \left\{\partial
G_0^{-1}/\partial k_b, s_i\right\}_+$ with $b = \mu, \nu $ or
$\lambda$ and $i =\alpha, \beta$ or $\gamma$ as appropriate. The
space-time indices $\mu, \nu$ and $\lambda$ take values $t$,$x$ and
$y$ whereas the spin indices $\alpha, \beta$ and $\gamma$ take
values $x$, $y$ and $z$. The straight lines denote the Green
function $G_0$ and the wavy lines denote the total SU(2) gauge
fields $\Omega^{\rm total}$}. \label{hopdia}
\end{figure}
To this end, we now integrate out the fermions fields from $S$ and
carry out a gradient expansion of the resultant effective action in
${C_{\mu}^{\rm total}}$ as done in Refs.\ \onlinecite
{ks3,stone1,volovik1}. We shall concentrate here only on the terms
in the effective action which contributes to the spin-Hall current.
These diagrams are shown in Fig.\ \ref{hopdia}. After some
straightforward, but tedious algebra, one obtains the effective
action which has the same form as that obtained for $^3$ He-A in
Ref.\ \onlinecite{volovik1}
\begin{eqnarray}
S_{\rm eff} &=& S_{\rm int} + S_{\rm ext} + S_{\rm coupling},
\label{ac5} \\
S_{\rm int} &=& \frac{N}{16 \pi} \epsilon_{\mu \nu \lambda}  \int
d^2r dt \, \Omega_{\nu}^{z\,\rm int} \partial_{\lambda}
\Omega^{z\,\rm
int}_{\mu},  \label{efint} \\
S_{\rm ext} &=& \frac{N}{32 \pi} \epsilon_{\mu \nu \lambda} \int
d^2r dt \, \Big [ {\bf \Omega}_{\mu}^{\rm ext} \cdot {\bf f} _{\nu
\lambda}^{\rm ext} \nonumber\\
&& -\frac{1}{3} {\bf \Omega}_{\mu}^{\rm ext}\cdot \left( {\bf
\Omega}_{\nu}^{\rm ext}\times {\bf \Omega}_{\lambda}^{\rm
ext}\right) \Big], \label{efext} \\
S_{\rm coupling} &=& \frac{1}{16 \pi} \int d^2r dt \,\Big[ N
\epsilon_{\mu \nu \lambda} \nonumber\\
&& \times \left( \partial_{\mu} {\bf d} - {\bf \Omega}_{\mu}^{\rm
ext} \times {\bf d} \right)\cdot \left( {\bf f}_{\nu \lambda}^{\rm
ext} \times {\bf d} \right) \nonumber\\
&& + 2 P  \left( \partial_{t} {\bf d} - {\bf \Omega}_{t}^{\rm ext}
\times {\bf d} \right)\cdot \left( {\bf f}_{xy}^{\rm ext} \times
{\bf d} \right)\Big]. \label{scoup}
\end{eqnarray}
Here ${\bf f} _{\mu \nu}^{\rm ext} = \partial_{\mu} {\bf
\Omega}_{\nu}^{\rm ext} - \partial_{\nu} {\bf \Omega}_{\mu}^{\rm
ext} -  {\bf \Omega}_{\mu}^{\rm ext} \times {\bf \Omega}_{\nu}^{\rm
ext}$ is the field-strength corresponding to the external SU(2)
fields, $\epsilon_{\mu \nu \lambda}$ is the antisymmetric tensor and
the coefficients $N$ and $P$ can be written in terms of the Green
function $G_0(k;\Delta_0)$ as \cite{volovik1,stone1}
\begin{eqnarray}
N &=& 2 \pi {\rm Tr} \Big[ G_0(k;\Delta_0) \frac{\partial
G_0^{-1}(k,\Delta_0)}{\partial k_x} G_0(k;\Delta_0) \nonumber\\
&& \times \frac{\partial G_0^{-1}(k,\Delta_0)}{\partial k_y}
G_0(k;\Delta_0) \frac{\partial G_0^{-1}(k,\Delta_0)}{\partial \omega
}\Big], \label{topin} \\
P &=& 2 \pi {\rm Tr} \Big[ G_0(k;\Delta_0) \frac{\partial
G_0^{-1}(k;\Delta_0)}{\partial \omega} G_0(k;\Delta_0) \nonumber\\
&& \times \frac{\partial G_0^{-1}(k;0)}{\partial k_x}
G_0(k;-\Delta_0) \frac{\partial G_0^{-1}(k,0)}{\partial k_y }\Big],
\label{notopin}
\end{eqnarray}
where ${\rm Tr}$ denotes matrix traces and sum over all frequencies
and momenta. Notice that $N$ is precisely the topological invariant
that characterizes the SU(2) Hopf term as shown in Refs.\
\onlinecite{stone1,volovik1}. For chiral triplet $p$-wave
superconductors and $^3$He-A, it is well-known that $N=1$. The
coefficient $P$, however, is not a topological invariant and its
numerical value depends on the details of the system. Evaluating the
integrals and matrix traces in Eq.\ \ref{notopin}, one finds that
$P=1$ as long as we restrict ourselves to clean systems and $T=0$.
In this work, we shall restrict ourselves within this domain. We
also note that $N$ is exactly quantized only if there are no nodes
in the superconducting gap; however, the deviation of $N$ from its
quantized value is expected to be vanishingly small in ruthenates in
the presence of the line of nodes, since the gap vanishes only along
a line of the entire three dimensional Fermi surface. We shall
ignore such deviation in the rest of this section.

In the absence of external Zeeman field, we are left with only the
internal fields ${\bf \Omega}_{\mu}^{\rm int}$. In this case, the
effective action $S_{\rm eff} =S_{\rm int}$ (Eqs.\
\ref{ac5},\ref{efint}) reduces to SU(2) Hopf term derived in Refs.\
\onlinecite{volovik1,stone1}. The contribution to the spin-current,
on the other hand, comes from $S_{\rm coupling}$ and $S_{\rm ext}$,
and thus requires the full effective action in the presence of a
finite Zeeman field. Using Eqs.\ \ref{scu1}, \ref{efext}, and
\ref{scoup}, we obtain the spin-Hall current for arbitrary
configuration of the ${\bf d}({\bf r},t)$ and for any relative
orientation of the applied Zeeman field ${\bf B}$ and ${\bf d}$
\begin{eqnarray}
{\bf j}_i({\bf r},t) &=& \frac{1}{8\pi} \epsilon_{ij} \Bigg[
\partial_j \left\{P \gamma {\bf B}  - \frac{1}{2} (N+2P){\bf d}
\left( {\bf
d}\cdot \gamma {\bf B} \right) \right\} \nonumber\\
&& - 2N \, \partial_j \left( {\bf d}  \times \partial_t {\bf d}
\right) - \frac{N}{2} {\bf d} \, \partial_j \left( {\bf d} \cdot
\gamma {\bf B} \right) \Bigg], \label{scurr1}
\end{eqnarray}
where $i$ and $j$ takes values $x,y$, $\gamma=g \mu_B$, and
$\epsilon_{ij}$ is the antisymmetric tensor with $\epsilon_{xy}=1$.
Note that Eq.\ \ref{scurr1} has to be supplemented with the equation
governing the dynamics of the ${\bf d}$ fields in the presence of an
external magnetic field. This has been derived, in the presence of a
pinning term, in Ref.\ \onlinecite{kee1}. For the rest of this
section, we shall concentrate on the case where the pinning
potential (and hence the pinning frequency) is larger than all other
scales in the problem, so that ${\bf d}$ is fixed along the c-axis.

 First, let us consider the case when
${\bf B} \parallel {\bf d} = B {\hat z}$. In this case one obtains a
quantized spin-Hall bulk current from Eq.\ \ref{scurr1}
\begin{eqnarray}
\left(j_i^z\right)^{\parallel} &=& -\frac{N}{8\pi}
\epsilon_{ij}\partial_j (g \mu_B B) = -\frac{1}{8\pi}
\epsilon_{ij}\partial_j (g \mu_B B).\label{bsc1}
\end{eqnarray}
This result differs from the conclusion of Ref.\
\onlinecite{volovik1}, where the bulk spin current is claimed to
vanish for magnetic field along the ${\bf d}$ vector. To understand
the reason for this difference, let us consider a superconducting
sample occupying semi-infinite space $x\ge 0$ with a edge at $x=0$.
In this case, since magnetic field $B$ decays to zero deep inside
the superconductor, the bulk spin-current is given by
\begin{eqnarray}
(I_y^z)_{\rm bulk}^{\parallel}&=& \int_0^{\infty}
\left(j_y^z\right)^{\parallel} dx \nonumber\\
&=&  - \frac{1}{8\pi} (g \mu_B B) = - (I_{\parallel}^z)_{\rm
edge}^{\rm spin},\label{bsc2}
\end{eqnarray}
where $(I_{\parallel}^z)_{\rm edge}^{\rm spin}$ is the edge spin
current given by Eq.\ \ref{is1}. Thus the bulk spin current is {\it
equal and opposite} to the edge current. This has an interesting
consequence. Imagine that we place the semi-infinite sample with in
a solenoid coil centered around the edge of the sample at $x=0$
which produces a magnetic field along ${\bf d}$. In this case, the
solenoid will create a both a bulk and the edge spin current so that
total spin current $(I_{\parallel}^z)_{\rm edge}^{\rm spin} +
(I_y^z)_{\rm bulk}^{\parallel} =0$. This is due to the fact that the
solenoid creates a potential difference but no net electrochemical
potential difference in the sample. Consequently, any net transverse
spin-current must vanish. This effect is analogous to the well-known
absence of charge quantum Hall current when a semiconducting sample
is placed between the plates of a capacitor. Thus we conclude that,
contrary to the claim in the Ref.\ \onlinecite{volovik1}, the bulk
spin current does not vanish for ${\bf B}
\parallel {\bf d}$; however the bulk and the edge spin current do cancel
out each other, and the total spin-current vanishes. Also note that
in this case the orbital effect of the magnetic field, which we have
neglected so far, do not contribute to the spin-current since the
condensate do not carry any net spin.

In contrast, when the magnetic field is applied perpendicular to the
${\bf d}$ vector (we choose ${\bf B} = B {\hat z}$ and $ {\bf d}
\parallel {\hat y}$), the bulk spin-current becomes
\begin{eqnarray}
(I_{y}^z)_{\rm bulk}^{\perp} &=& \int_0^{\infty}
\left(j_y^z\right)^{\perp} dx \nonumber\\
&=& \frac{P}{8\pi} (g \mu_B B) =  \frac{1}{8\pi} (g \mu_B B).
\label{bsc3}
\end{eqnarray}
Thus we find that there is a net bulk spin-Hall current for ${\bf B}
\perp {\bf d}$ which is in accordance with the conclusion of Ref.\
\onlinecite{volovik1}. Notice that in this case, as we have seen in
the Sec.\ \ref{ruedge}, the edge spin current is already screened by
the Meissner current and hence it does not contribute to the net
spin current. By taking into account the generation of Meissner
current, we have qualitatively taken into account the orbital effect
of the applied magnetic field. A detailed numerical computation,
following the lines of Ref.\ \onlinecite{sigrist1}, is required to
address this issue on a more quantitative level and is beyond the
scope of the present work. Also, we note that $(I_{y}^z)_{\rm
bulk}^{\perp}$ depends on $P$ and not the SU(2) Hopf coefficient
$N$, as can be easily seen from Eq.\ \ref{scurr1}. Thus we conclude
that the bulk spin-Hall conductivity in a triplet chiral
superconductor (and also $^3$He-A) for ${\bf B} \perp {\bf d}$ is
not truly quantized, although for clean systems at $T=0$ where
$P=1$, it reaches its quantized value $1/8\pi$.

\section{Graphene}
\label{graphene}

The low energy Hamiltonian of graphene, in the presence of Rashba
coupling, is given by \cite{kane1}
\begin{eqnarray}
H &=& \int d^2 r \psi^{\dagger}({\bf r}) \Big[-i v_F \left( \tau_z
\sigma_x
\partial_x + \sigma_y
\partial_y\right) + \Delta_1 \tau_z \sigma_z s_z \nonumber\\
&& + \lambda_R \left( \tau_z s_y \sigma_x - s_x \sigma_y\right)\Big]
\psi({\bf
r}),\nonumber \\
\label{grham1}
\end{eqnarray}
where $v_F$ is the Fermi velocity, $\tau_i$, $\sigma_i$ and $s_i$
are the Pauli matrices in the $K-K'$, $A-B$ and spin spaces
respectively, $\Delta_1$ is the spin-orbit coupling gap, and
$\lambda_R$ denotes the Rashba coupling. Note that the spin-orbit
coupling gap does not break time reversal symmetry on the whole, but
does so for each individual spin species. Here $\psi$ is a $8$
component fermionic spinor field given by $\psi = \left( \psi_{A
\uparrow}^K, \psi_{B \uparrow}^K,\psi_{A \uparrow}^{K'},\psi_{B
\uparrow}^{K'},\psi_{A \downarrow}^K, \psi_{B \downarrow}^K, \psi_{A
\downarrow}^{K'}, \psi_{B \downarrow}^{K'}\right)$. In the rest of
this section, we shall analyze this Hamiltonian in the presence of
an weak external electromagnetic field, and also study its energy
spectrum in the presence of an armchair edge. Our derivation can in
principle to other models for spin-Hall effect which has similar
structure of the effective low-energy Hamiltonian \cite{rahul1}

\subsection{Bulk effective action}
\label{grbulk}

In this section, we derive a low energy effective action of graphene
in the presence of an external electromagnetic field and obtain the
expression of spin-current from that effective action. First, we
study the system in the absence of Rashba coupling. We begin with
the action
\begin{eqnarray}
S &=& \int d^2r d t   \psi^{\dagger}({\bf r},\tau) \Bigg [ i
\partial_{t}
-eA_0 - v_F \Big( \tau_z \sigma_x (-i\partial_x -eA_x) \nonumber\\
&& + \sigma_y (-i\partial_y -eA_y) \Big )- \Delta_1 \tau_z \sigma_z
{\bf s} \cdot {\bf d}({\bf r},t) \Bigg]\psi({\bf r},\tau),
\label{grac1}
\end{eqnarray}
where $A_{\mu}$ denotes the $U(1)$ gauge fields corresponding to an
external electric field $E_{i} = -\partial_t A_i - \partial_i A_0$.
Note that here we have introduced an unit vector field ${\bf d}({\bf
r},\tau)$. The Hamiltonian $H$ (Eq.\ \ref{grham1}) corresponds to
the configuration ${\bf d} =z$ and we shall set ${\bf d} =z$ at the
end of the calculation.

\begin{figure}
\rotatebox{0}{
\includegraphics*[width=\linewidth]{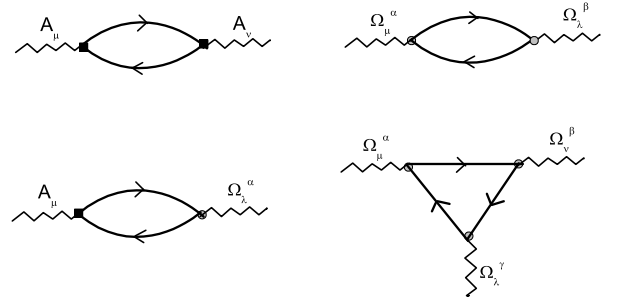}}
\caption{Feynman diagrams which contribute to the possible Hopf
terms effective action. The left panel contains diagram contributing
to $S^{(1)}$ and $S^{(3)}$ where as the right panel shows diagrams
contributing to $S^{(2)}$. The filled squares represent the vertices
$\partial G_0^{-1}/\partial k_b$ with $b = \mu$ or $\nu$, as
appropriate. The closed circles represent $ 1/2 \left\{
\partial G_0^{-1}/\partial k_b, s_i\right\}_+$ as in Fig.\ \ref{hopfdia},
whereas the straight lines denote the Green function $G_0$.}
\label{feyn}
\end{figure}

Next, as in Sec.\ \ref{rubulk}, we introduce a SU(2) rotation in the
spin-space $ \psi({\bf r},\tau) \rightarrow \psi'({\bf r},\tau) =
U^{\dagger} \psi({\bf r},\tau)$ where $U$ is a local SU(2) rotation
matrix which rotates the local ${\bf d}$ vector to $z$:
$U^{\dagger}\,{\bf s} \cdot {\bf d}\, U = s_z$. The partition
function can now be written in terms of the new spinor fields
$\psi'(k) \equiv \psi'({\bf k},\omega )$
\begin{eqnarray}
Z &=& \int {\mathcal D} \psi'^{\dagger} {\mathcal D} \psi'
{\mathcal D} C_{\mu}  {\mathcal D}A_{\mu}  \exp\left[i(S_0 + S_1)\right], \nonumber\\
S_0 &=& \int \frac{d^2 k d \omega}{(2\pi)^3}
\psi'^{\dagger}(k) G_0^{-1}(k)  \psi' (k), \label{s0} \\
S_1 &=& - \int \frac{d^2
k d^2p d\omega d p_0 }{(2\pi)^6} \psi'^{\dagger}(k+p) \nonumber\\
&& \times \frac{1}{2} \left \{
\partial_{k_{\mu}} G_0^{-1}(k), eA_{\mu}(p) + C_{\mu}(p)
\right\}_+ \psi'(k), \label{pert}
\end{eqnarray}
where the fields $C_{\mu} = -iU^{\dagger} \partial_{\mu} U = 1/2\,
{\bf s} \cdot {\bf \Omega}_{\mu}$ are the SU(2) gauge fields, the
configuration ${\bf d} = z$ corresponds to $C_{\mu}=0$ and we have
used the same notations as in Sec.\ \ref{rubulk}. Here the Green
function $G_0(k; \Delta_1)$ is given by
\begin{eqnarray}
G_0(k;\Delta_1) &=& \frac{\omega + v_F(\tau_3 k_x \sigma_x + k_y
\sigma_y) + \Delta_1 \tau_z \sigma_z s_z }{\omega^2 - E_k^2 +i
\eta},
\label{green1} \nonumber\\
E_k &=& \sqrt {v_F^2 (k_x^2 +k_y^2) + \Delta_1^2}. \label{en1}
\end{eqnarray}
Note that the $\psi'$ fields see a gap term $\Delta_1 \sigma_z
\tau_z s_z$ and therefore describes the same action as the one
corresponding to $H$ (Eq.\ \ref{grham1}) in the absence of the
$C_{\mu}$ fields. The advantage of introducing the ${\bf d}$ fields
now become clear, since we find that the spin-current density
$j_{\mu}^{\alpha} = \left<\psi'^{\dagger} \frac{1}{2} \left
\{\partial_{k_{\mu}} G_0^{-1}(k), s^{\alpha} \right\}_+ \psi'
\right>$ can be obtained as
\begin{eqnarray}
j_{\mu}^{\alpha} &=& \frac{\partial S_{\rm
eff}[C_{\mu},A_{\mu}]}{\partial \Omega_{\mu}^{\alpha}} \Bigg
|_{\Omega_{\mu}^{\alpha}=0}, \label{scurrent}
\end{eqnarray}
where $S_{\rm eff}$ is the effective action obtained by integrating
out the Fermion fields. Here the $\mu =t$ component of the
spin-current is simply the spin-density $ {\bf j}_{t} = \left<
\psi'^{\dagger} {\bf s} \psi' \right>$.

To obtain the effective action, we now integrate out the fermionic
fields and compute the effective action for $A_{\mu}$ and $C_{\mu}$
fields. Since for slowly varying ${\bf d}$ field configuration
$C_{\mu}$ is small, we can carry out a gradient expansion in powers
of the fields $A_{\mu}$ and $C_{\mu}$ and their derivatives in a
straightforward manner. The relevant terms in the effective action
which contributes to the Hopf terms comes from the first derivative
of the polarization bubble and triangle diagram shown in Fig.\
\ref{feyn}. These terms are given by
\begin{eqnarray}
S_{\rm eff}\left[A_{\mu}, C_{\mu}\right] &=& S^{(1)} + S^{(2)}
+ S^{(3)}, \\
S^{(1)} &=& a_{\mu \nu \lambda} \int d^2r dt\, A_{\nu}
\,\partial_{\lambda}\, A_{\mu}, \label{schern} \\
S^{(2)} &=& b_{\mu \nu \lambda}^{\alpha \beta}  \int d^2r dt
\,\Omega_{\nu}^{\alpha}\,
\partial_{\lambda} \,\Omega_{\mu}^{\beta} \nonumber\\
&& + c_{\mu \nu \lambda}^{\alpha \beta \gamma} \int d^2r dt
\,\Omega_{\mu}^{\alpha}
\,\Omega_{\nu}^{\beta}\, \Omega_{\lambda}^{\gamma},  \label{shopf} \\
S^{(3)} &=& d_{\mu \nu \lambda}^{\alpha} \int d^2r dt \,A_{\nu}
\,\partial_{\lambda} \,\Omega_{\mu}^{\alpha}, \label{scrossed}
\end{eqnarray}
where we have used the definition $C_{\mu} = 1/2 s_{\alpha}
\Omega_{\mu}^{\alpha}$, and all repeated indices are summed over.
The coefficients of the terms in the effective action can be
expressed in terms of Green functions as
\begin{eqnarray}
a_{\mu \nu \lambda} &=& -\frac{e^2}{2} {\rm Tr} \left(
\frac{\partial G_0^{-1}}{\partial k_{\mu}} G_0(k;\Delta_1)
\frac{\partial G_0^{-1}}{\partial k_{\nu}}  \right.
\nonumber\\
&& \left. \times G_0(k;\Delta_1) \frac{\partial G_0^{-1}}{\partial
k_{\lambda}}
G_0(k;\Delta_1) \right), \label{coeffchern} \\
b_{\mu \nu \lambda}^{\alpha \beta} &=& - \frac{1}{8} {\rm Tr} \left(
\left\{ \frac{\partial G_0^{-1}}{\partial k_{\mu}},s_{\alpha}
\right\}_+ G_0(k;\Delta_1) \left\{\frac{\partial G_0^{-1}}{\partial
k_{\nu}},
s_{\beta} \right\}_+ \right. \nonumber\\
&& \left. \times  G_0(k;\Delta_1) \frac{\partial G_0^{-1}}{\partial
k_{\lambda}}
G_0(k;\Delta_1) \right), \label{coeffh2} \\
c_{\mu \nu \lambda}^{\alpha \beta \gamma} &=& \frac{i}{8} {\rm Tr}
\left( \left\{ \frac{\partial G_0^{-1}}{\partial k_{\mu}},s_{\alpha}
\right\}_+ G_0(k;\Delta_1) \left\{\frac{\partial G_0^{-1}}{\partial
k_{\nu}},
s_{\beta} \right\}_+   \right. \nonumber\\
&& \left. \times G_0(k;\Delta_1)\left\{\frac{\partial
G_0^{-1}}{\partial k_{\lambda}}
,s_{\gamma}  \right\}_+ G_0(k;\Delta_1) \right), \label{coeffh3} \\
d_{\mu \nu \lambda}^{\alpha} &=& - \frac{e}{4} {\rm Tr} \left(
\left\{ \frac{\partial G_0^{-1}}{\partial k_{\mu}},s_{\alpha}
\right\}_+ G_0(k;\Delta_1) \frac{\partial G_0^{-1}}{\partial
k_{\nu}}
\right. \nonumber\\
&& \left. \times G_0(k;\Delta_1) \frac{\partial G_0^{-1}}{\partial
k_{\lambda}} G_0(k;\Delta_1) \right), \label{coeffh4}
\end{eqnarray}
where ${\rm Tr}$ denotes trace over all Pauli matrices and
integration over frequencies and momenta. We now need to evaluate
these coefficients and we begin with $b_{\mu \nu \lambda}^{\alpha
\beta}$ and $c_{\mu \nu \lambda}^{\alpha \beta \gamma}$. We note
that for $b_{\mu \nu \lambda}^{\alpha \beta}$, all the coefficients
with $\alpha \ne z$ and $\beta=z$ or vice versa vanish when we take
trace over Pauli matrices in spin-space. The same line of reasoning
shows that only the coefficients $ c_{\mu \nu \lambda}^{\alpha
\alpha z}$ or $ c_{\mu \nu \lambda}^{xyz}$ can be non-zero. Further
it is easy to see that only $d_{\mu \nu \lambda}^{z}$ has
non-vanishing trace. Thus the only non-vanishing coefficients can be
written as
\begin{eqnarray}
b_{\mu \nu \lambda}^{xx/yy} &=& - \frac{1}{2} {\rm Tr} \left(
\frac{\partial G_0^{-1}}{\partial k_{\mu}} G_0(k;-\Delta_1)
\frac{\partial
G_0^{-1}}{\partial k_{\nu}}  G_0(k;\Delta_1) \right. \nonumber\\
&& \left. \times \frac{\partial G_0^{-1}}{\partial k_{\lambda}}
G_0(k;\Delta_1) \right) = c_{\mu \nu \lambda}^{xyz}, \\ \label{hc1}
b_{\mu \nu \lambda}^{xy} &=& - \frac{i}{2}  {\rm Tr} \left( s_z
\frac{\partial G_0^{-1}}{\partial k_{\mu}} G_0(k;-\Delta_1)
\frac{\partial
G_0^{-1}}{\partial k_{\nu}}  G_0(k;\Delta_1) \right. \nonumber\\
&& \left. \times \frac{\partial G_0^{-1}}{\partial k_{\lambda}}
G_0(k;\Delta_1) \right) \nonumber\\
&=& c_{\mu \nu \lambda}^{xxz} = c_{\mu \nu \lambda}^{yyz} = - b_{\mu
\nu \lambda}^{yx}, \label{hc2} \\
b_{\mu \nu \lambda}^{zz} &=& - \frac{1}{2} {\rm Tr} \left(
\frac{\partial G_0^{-1}}{\partial k_{\mu}} G_0(k;\Delta_1)
\frac{\partial
G_0^{-1}}{\partial k_{\nu}}  G_0(k;\Delta_1) \right. \nonumber\\
&& \left. \times \frac{\partial G_0^{-1}}{\partial k_{\lambda}}
G_0(k;\Delta_1) \right) = a_{\mu \nu \lambda}/e^2, \label{hc3}\\
d_{\mu \nu \lambda}^{z} &=& - \frac{e}{2} {\rm Tr} \left(s_z
\frac{\partial G_0^{-1}}{\partial k_{\mu}} G_0(k;\Delta_1)
\frac{\partial
G_0^{-1}}{\partial k_{\nu}}  G_0(k;\Delta_1) \right. \nonumber\\
&& \left. \times \frac{\partial G_0^{-1}}{\partial k_{\lambda}}
G_0(k;\Delta_1) \right) = c_{\mu \nu \lambda}^{zzz}.
\end{eqnarray}
It turns out \cite{ks3,volovik1} that coefficients $b_{\mu \nu
\lambda}^{\alpha \beta}$ also satisfy the identity $b_{\mu \nu
\lambda}^{\alpha \beta}= -b_{\nu \mu \lambda}^{\alpha \beta}$. Using
this, we find that the contribution of terms proportional to $b_{\mu
\nu \lambda}^{xy}$ in the effective action leads to a total
derivative which we ignore and those due to $c_{\mu \nu
\lambda}^{\alpha \alpha z}$ vanish identically. Further one can use
the fact that the fields ${\bf \Omega}_{\mu}$ are pure gauge fields
and hence satisfy the identity
\begin{eqnarray}
\partial_{\mu}{\bf \Omega}_{\nu} - \partial_{\nu}{\bf \Omega}_{\mu}
- {\bf \Omega}_{\mu} \times {\bf \Omega}_{\nu} &=& 0. \label{fint}
\end{eqnarray}
Using Eqs.\ \ref{fint} and \ref{hc1}, it is easy to show that all
the terms proportional to $b_{\mu \nu \lambda}^{xx/yy}$ vanish. Thus
finally we are left with the effective action
\begin{eqnarray}
S^{(1)}&=&  a_{\mu\nu\lambda} \int d^2r dt\, A_{\nu}
\,\partial_{\lambda}\, A_{\mu}, \label{schern1} \\
S^{(2)} &=& \frac{a_{\mu \nu \lambda}}{e^2} \int d^2r dt
\,\Omega_{\nu}^{z}\,
\partial_{\lambda} \,\Omega_{\mu}^{z}, \label{shopf1} \\
S^{(3)} &=& d_{\mu \nu \lambda}^{z} \int d^2r dt \,A_{\nu}
\,\partial_{\lambda} \,\Omega_{\mu}^{z}. \label{scrossed1}
\end{eqnarray}
The next task is to evaluate the coefficient $a_{\mu \nu \lambda}$,
which can be done by straightforward evaluation of integrals. But
before resorting to algebraic manipulation, it is easier to note
that one needs $a_{\mu \nu \lambda} = e^2 \epsilon_{\mu \nu \lambda}
a$ for electromagnetic gauge invariance of $S^{(1)}$, where $a$ is
given by
\begin{eqnarray}
a &=& -\frac{1}{2} {\rm Tr} \left( \frac{\partial G_0^{-1}}{\partial
k_{x}} G_0(k;\Delta_1) \frac{\partial
G_0^{-1}}{\partial k_{y}}  G_0(k;\Delta_1) \right. \nonumber\\
&& \left. \times \frac{\partial G_0^{-1}}{\partial \omega}
G_0(k;\Delta_1) \right). \label{aterm}
\end{eqnarray}
Further, we find that the contribution to $a$ in Eq.\ \ref{aterm} is
a sum of contributions from spin $\uparrow$ and $\downarrow$
sectors: $a= a_{\uparrow} + a_{\downarrow}$. Whereas each of these
two contributions can be finite since the Hamiltonian (Eq\
\ref{grham1}) breaks time reversal symmetry for each spin species,
their sum must vanish since the total Hamiltonian for both the spin
species is time reversal invariant. Thus we conclude that $a=0$ and
hence $b_{\mu \nu \lambda}^{zz}=0$, so that graphene does not
support charge Hall effect in the absence of an external magnetic
field and also does not have a pure SU(2) Hopf term. However, the
same argument tells us that $d_{\mu \nu \lambda}^z = -e
\epsilon_{\mu \nu \lambda} d \ne 0$, since $ d= a_{\uparrow} -
a_{\downarrow} \ne 0$. Indeed a straightforward evaluation of $a$
shows $ a_{\uparrow} = - a_{\downarrow} = - 1/2\pi$. Hence we
conclude that the effective action of graphene supports a crossed
Hopf term
\begin{eqnarray}
S^{(3)} &=& \frac{e}{2\pi} \epsilon_{\mu \nu \lambda} \int d^2r dt
\,\Omega_{\mu}^z \, F_{\nu \lambda}, \label{chopf}
\end{eqnarray}
where $F_{\nu \lambda} =\partial_{\nu} A_{\lambda} -
\partial_{\lambda} A_{\nu}$ is the electromagnetic field tensor.
This leads to a quantized Hall spin-current (Eq.\ \ref{scurrent})
\begin{eqnarray}
j_{i}^z = \epsilon_{ij} \frac{e}{2\pi} E_j,
\end{eqnarray}
where ${\bf E} = -\nabla A_0 - \partial_t {\bf A}$ is the applied
electric field.

Next we introduce the Rashba term in Eq.\ \ref{grham1}. We shall
restrict the analysis to the case of $\lambda_R /\Delta_1 \ll 1$ and
assume that the strength of the Rashba term is not sufficient to
destroy the spin-Hall phase.  In the presence of such a term, the
inverse of the Green function in $S_0$ (Eq.\ \ref{s0}) becomes
\begin{eqnarray}
G_0^{-1\,R} &=& \omega -v_F(\tau_3 k_x \sigma_x + k_y \sigma_y) -
\Delta_1 \tau_z \sigma_z s_z \nonumber\\
&& -\lambda_R \left( \tau_z s_y \sigma_x - s_x \sigma_y\right).
\label{greenrashba}
\end{eqnarray}
One can now repeat the same analysis as described above with
$G_0(k;\Delta_1)$ replaced by $G_0^R(k;\Delta_1,\lambda_R)$. Since
$\partial G_0^{-1\,R}/\partial k_{\mu}$ is identical to $\partial
G_0^{-1}/\partial k_{\mu}$, one finds that all the arguments
regarding the coefficients $a_{\mu \nu \lambda},b_{\mu \nu
\lambda}^{\alpha \beta}, c_{\mu \nu \lambda}^{\alpha \beta \gamma}$,
and $d_{\mu \nu \lambda}^{\alpha}$ remain the same and, in the end,
we are left with
\begin{eqnarray}
S^{(3)} &=& d^R \epsilon_{\mu \nu \lambda} \int d^2 dt \,
\Omega_{\mu}^z \, F_{\nu \lambda},\nonumber\\
d^{R} &=& \frac{e}{2} {\rm Tr} \left(s_z \frac{\partial
G_0^{-1\,R}}{\partial k_{x}} G_0^R(k;\Delta_1,\lambda_R)
\frac{\partial
G_0^{-1\,R}}{\partial k_{y}}   \right. \nonumber\\
&& \left. \times G_0^R(k;\Delta_1,\lambda_R) \frac{\partial
G_0^{-1\,R}}{\partial \omega} G_0^R(k;\Delta_1,\lambda_R) \right),
\nonumber\\
&=& \frac{e}{2\pi} \left(1- \frac{\lambda_R^2}{6\Delta_1^2 }+
...\right),
\end{eqnarray}
where the ellipsis represent terms which are higher order in
$\lambda_R/\Delta_1$, and we have obtained the last line by
explicitly evaluating the matrix traces and frequency and momentum
integrals in the expression of $d^{R}$.  Thus we find that the
spin-Hall conductivity deviates from its quantized value and equals
$ e(1 - \lambda_R^2/(6 \Delta_1^2))/2\pi$ for $\lambda_R/\Delta_1
\ll 1$.

\subsection{Edge states}
\label{gredge}

The edge states in graphene in the absence of any external magnetic
field can be analytically obtained from the Dirac Hamiltonian $H$
(Eq.\ \ref{grham1}) for the armchair edge. Throughout this section,
we shall restrict ourselves to the case $\lambda_R=0$. The geometry
we study here is that of a semi-infinite sample occupying $x>0$ with
an armchair edge at $x=0$. For such an edge, the boundary condition
demands both $\psi_A$ and $\psi_B$ to vanish at $x=0$. In what
follows, we construct an analytical solution for the low-energy
subgap localized energy states of the graphene hamiltonian $H$ (Eq.\
\ref{grham1}) which respects the boundary condition $\psi_A (x=0)=0$
and $\psi_B(x=0)=0$. Notice that in doing so by starting from Dirac
equation, we ignore the lattice effects at the edge, which usually
puts a momentum cut-off $k_c$ above which the present description
breaks down. The estimate of $k_c$ can not be reliably done without
going into numerical analysis of the lattice Hamiltonian
\cite{brey1,kane1}.

To construct such a solution, we note that the only way to achieve
the above-mentioned boundary condition is to superpose
wave-functions in the $K$ and $K'$ points. We therefore try a
normalized wavefunction of the form
\begin{equation}
\psi_{\rm edge}(x,k_y) = e^{ik_y y -\kappa x} \sqrt{\kappa} \left[
{u_{A\,s}^{K}\choose u_{B \,s}^{K}} e^{i {\bf K}\cdot {\bf r}} -
{u_{A\,s}^{K'}\choose u_{B \,s}^{K'}} e^{i {\bf K'}\cdot {\bf r}}
\right]. \label{wave}
\end{equation}
Here $s=\uparrow,\downarrow$ denotes the spin of the electrons,
${\bf K}$ and ${\bf K'}$ are the wavevectors of the $K$ and $K'$
points in Brillouin zone of graphene, $\kappa^{-1}$ denotes
localization length of the edge states, and $k_y$ denotes momenta
parallel to the edge. The normalization condition $\sum_s \int dx
\left|\psi_{\rm edge}(x)\right|^2 =1$ of the wave functions imply
$\sum_s \left|u_{A\,s}^{K(K')}\right|^2 +
\left|u_{B\,s}^{K(K')}\right|^2=1$ and the boundary condition
$\psi_{\rm edge}(x)=0$ necessitates the condition
\begin{eqnarray}
\frac{u_{A\,s}^{K}}{u_{B \,s}^{K}} &=& \frac{u_{A\,s}^{K'}}{u_{B
\,s}^{K'}}. \label{bc}
\end{eqnarray}

From Eqs.\ \ref{grham1} and \ref{wave}, it is easy to see that the
wavefunctions $\left(u_{A\,s}^{K},u_{B\,s}^{K}\right)$ and $
\left(u_{A\,s}^{K'},u_{B\,s}^{K'}\right)$ satisfy Schrodinger
equations
\begin{eqnarray}
  &&\left(\begin{array}{cc}
    {\rm sgn}(s)\Delta_1 & i v_F(\kappa-k_y) \\
    i v_F(\kappa + k_y) & -{\rm sgn}(s)\Delta_1
    \end{array}\right)
    {u_{A\,s}^K \choose u_{B\,s}^K}
    =E {u_{A\,s}^K \choose u_{B\,s}^K},
    \nonumber\\
&&\left(\begin{array}{cc}
    -{\rm sgn}(s)\Delta_1 & -i v_F(\kappa+k_y) \\
    -i v_F(\kappa - k_y) & {\rm sgn}(s)\Delta_1
    \end{array}\right)
    {u_{A\,s}^{K'} \choose u_{B\,s}^{K'}}
    =E {u_{A\,s}^{K'} \choose u_{B\,s}^{K'}}, \label{sch}
    \nonumber\\
\end{eqnarray}
where ${\rm sgn}(s)=+(-)$ for $s=\uparrow(\downarrow)$. Solving the
Schrodinger's equations, one obtains
\begin{eqnarray}
\frac{u_{A\,s}^{K}}{u_{B \,s}^{K}} &=&  \frac{E-{\rm sgn}(s)
\sqrt{E^2+v_F^2(\kappa^2-k_y^2)}}{i v_F (\kappa - k_y)}, \label{sch1} \\
\frac{u_{A\,s}^{K'}}{u_{B \,s}^{K'}} &=&  -\frac{E+{\rm sgn}(s)
\sqrt{E^2+v_F^2(\kappa^2-k_y^2)}}{i v_F (\kappa + k_y)}, \label{sch2} \\
v_F \kappa &=& \sqrt{ \Delta_1^2 -E^2 + v_F^2 k_y^2}. \label{sch3}
\end{eqnarray}
\begin{figure}
\rotatebox{0}{
\includegraphics*[width=\linewidth]{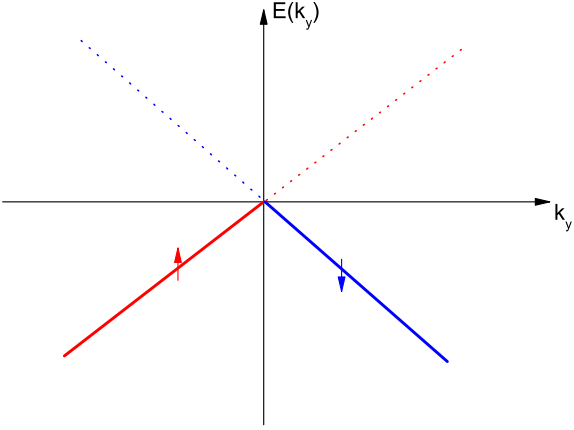}}
\caption{Edge states near an armchair edge in graphene. The solid
and dotted lines show occupied and empty spin-up(red) and spin-down
(blue) edge states. An equal occupation of the spin-up and spin-down
edge states result in cancelation of charge current.}
\label{spinedge}
\end{figure}
The boundary condition $\psi(x=0)=0$ (Eq.\ \ref{bc}) now yields the
condition
\begin{eqnarray}
\frac{E-{\rm sgn}(s) \sqrt{E^2+v_F^2(\kappa^2-k_y^2)}}{E+{\rm
sgn}(s) \sqrt{E^2+v_F^2(\kappa^2-k_y^2)}} &=& \frac {k_y
-\kappa}{k_y + \kappa}. \label{eneq}
\end{eqnarray}

Together with Eq.\ \ref{sch3}, this gives us localized states at the
edge with energy $E(k_y)$ and localization length $\kappa^{-1}$
\begin{eqnarray}
E(k_y) &=&  {\rm sgn}(s) v_F k_y, \quad \kappa^{-1} =
\frac{v_F}{\Delta_1}, \label{edgesol} \\
 \psi_{\rm edge} (x,k_y) &=&  \frac{\sqrt{\kappa}}{2} e^{i k_y y - \kappa x}
 e^{-i {\rm sgn}(s) \pi/4} \nonumber\\
 && \times \left[ {1\choose i {\rm sgn}(s)} e^{i {\bf K}\cdot {\bf r}}
 - {1\choose i {\rm sgn}(s)} e^{i
{\bf K'}\cdot {\bf r}}\right]. \label{wavefn} \nonumber\\
\end{eqnarray}
Notice that the analysis of the Dirac Hamiltonian predicts linear
dispersion of edge states for all $k_y$. This is clearly an artifact
throwing out the lattice and so we would expect the results to break
down at momenta $k_y^{\rm max} = k_c$, as commented earlier.

The edge states, in contrast to their chiral superconducting
counterparts, have {\it opposite} group velocities for spin-up and
spin-down electrons as shown in Fig.\ \ref{spinedge}. Further, in a
transverse momentum state $k_y$,  the edge states carry a charge
current
\begin{eqnarray}
j_y (k_y) &=&  e v_F n(k_y) \left< \psi_{\rm edge} \right| \sigma_y
\left| \psi_{\rm edge}  \right> = e {\rm sgn}(s) v_F n(k_y),
\label{ech1} \nonumber\\
\end{eqnarray}
where we have neglected the small $O(\kappa/|K_x-K'_x|)$
contributions. Note that the charge current is carried with the
group-velocity $v_g = {\rm sgn}(s) v_F$ and thus has opposite
direction for opposite spins. Hence these edge states do not carry a
net charge current for equal occupation of spin-up and spin-down
electron states, but carry a net spin-current circulating along the
edge. Note that an estimation of the total spin-current carried by
the edge requires a knowledge of the states at large $k_y$, and
hence can not be reliably done within this formalism. However, we
would like to stress that the low-energy linear dispersion of the
edge states with opposite group velocities for spin-up and spin-down
electrons agrees well with the numerical calculations of Refs.\
\cite{kane1,haldane1}.

\section{Discussion}
\label{conclusion}

The most direct way of detecting spin/charge current at the edge is
to create a population imbalance of the spin-up and spin-down edge
states. Such experiments has been performed for quantum Hall samples
in Ref.\ \onlinecite{ashoori1} and suggested for ruthenates in Ref.\
\onlinecite{ks2}. The idea is to apply a magnetic pulse ${\bf B}(t)$
to the edge of a chiral triplet superconductor with ${\bf B}
\parallel {\bf d}$. This creates a imbalance in the population of
the spin-up and spin-down edge states leading to local magnetization
which travels with the group velocity $v_e$. The magnetic moment due
to this spin imbalance can be detected by a squid magnetometer as
pointed out in Ref.\ \onlinecite{ks2}. The spin structure of the
edge states can also be verified since a similar experiment with
${\bf B} \perp {\bf d}$ will produce a null result. This feature
distinguishes the triplet chiral superconductors from their singlet
counterparts where the direction of the applied pulsed magnetic
field does not matter.

A similar experiment can be performed to verify the edge state
picture for graphene. Let us consider creating a population
imbalance for sub-gap edge states in graphene by applying a magnetic
field pulse $B$ along $z$. In the absence of spin-flip scattering
and for $B \le \Delta_1$ where the states in the bulk are gapped,
such an imbalance would persist and lead to a circulating quantized
charge current. The spin-flip scattering rate in graphene is
expected to be small since the Rashba coupling $\lambda_R \simeq
0.5$mK is small compared to the spin-orbit gap $\Delta_1 \simeq
1.2$K, and thus we expect the population imbalance and hence the
charge current to persist for a sufficiently long time
\cite{abanin1}. The magnitude of this charge current depends only on
the population imbalance between up and down spin low-energy subgap
electrons and hence, unlike the edge spin-current, can be reliably
estimated from the edge state picture developed in Sec.\
\ref{gredge}. Using Eq.\ \ref{ech1}, we obtain for the magnetic
pulse along $z$
\begin{eqnarray}
I_{\rm edge}^{\rm charge}  &=& e v_F \sum_{k_y} \left(
n_{\uparrow}(k_y) - n_{\downarrow}(k_y) \right) = \frac{2e}{h} g
\mu_B B \label{c1}.
\end{eqnarray}
This quantized charge-current for $g\mu_B B = \Delta_1 = 1.2 K$
leads to a net current of $I_{\rm edge}^{\rm charge} \simeq 8$nA
which can be easily detected by a setup similar to that proposed in
Ref.\ \onlinecite{ashoori1}. One of the central feature of the edge
states in graphene is that the direction of this charge edge current
depends on the direction of the applied magnetic pulse. This feature
can also be detected by a time-of flight measurement analogous to
those suggested in Refs. \onlinecite{ashoori1,ks2}. Alternatively,
one can also selectively populate the up or down-spin edge states by
spin-polarized tunneling into the edge states. The direction of the
edge current measured in such a tunneling experiment would  depend
on the polarization of the tunneling electrons and would serve as a
definitive proof of the spin-dependent velocity of the edge states.

In conclusion, we have studied the bulk and edge properties of TRS
broken and TRS invariant systems such as triplet chiral
superconductors and graphene. We have shown that both the structure
of the edge states and the topological terms in the bulk effective
action of these systems depend crucially on whether TRS is preserved
or not. For TRS broken systems such as triplet chiral
superconductor, the edge states carry a net charge current and the
bulk action contains a SU(2) Hopf leading to a spin-Hall current in
response to the gradient of applied Zeeman magnetic field. We have
also shown that both the edge and the bulk spin-Hall current of
triplet superconductors, in response to an applied Zeeman field
${\bf B}$, depend crucially on whether ${\bf B}\parallel {\bf d}$ or
${\bf B} \perp {\bf d}$. This is a consequence of broken
spin-rotational symmetry in these systems and serves as a
distinguishing feature of triplet chiral superconductors from their
singlet counterparts. In contrast, for TRS invariant systems such as
graphene, the edge states carry a net spin current while the bulk
action contains a crossed Hopf term leading to a spin-Hall current
in response to an external electric field. We have also suggested
experiments to verify our some of our results.

KS thanks V.M. Yakovenko and H-J. Kwon for earlier collaborations on
related projects. RR thanks M. Stone and E. Fradkin for useful
discussions and University of Illinois Research Board for support.


\begin{thebibliography}{99}

\bibitem{leggett1} A.J. Leggett, Rev. Mod. Phys. {\bf 47}, 331
(1975).

\bibitem{maeno1} A.P. Mackenzie and Y. Maeono, Rev. Mod. Phys. {\bf 75}, 657
(2003).

\bibitem{senthil1} T. Senthil, J. B. Marston, and M. P. A.
Fisher, Phys. Rev. B {\bf 60}, 4245 (1999)

\bibitem{ks1} K. Sengupta, H-J Kwon, and V.M. Yakovenko,
Phys. Rev. B {\bf 65}, 104504 (2002)

\bibitem{sigrist1} A. Furusaki, M. Matsumoto, and M. Sigrist, Phys. Rev. B {\bf 64},
054514 (2001).

\bibitem{matsumoto1} M. Matsumoto and R. Heeb, Phys. Rev. B {\bf 65}, 014504 (2002)

\bibitem{goryo1} J. Goryo, J. Phys. Soc. Jap. {\bf 69}, 3501-3504
(2000).

\bibitem{volovik1} G.E. Volovik and V. M. Yakovenko, J. Phys. Cond.
Mat. {\bf 1}, 5263 (1989); G.E. Volovik, A, Solov'ev, and V.M.
Yakovenko, JETP Lett. {\bf 49}, 65 (1989); For a review, see G.E.
Volovik, {\it Exotic Properties of Superfluid $^3$He} (World
Scientific, New Jersey (1993) ).

\bibitem {stone1} M. Stone and R. Roy, Phys. Rev. B {\bf 69}, 184511
(2004).

\bibitem{murakami1} S. Murakani, N. Nagaosa and S-C. Zhang, Science {\bf 301},
1348 (2003); J. Sinova {\it et al}, PRL {\bf 92}, 126603 (2004).

\bibitem {kane1} C. L. Kane and E. J. Mele, Phys. Rev. Lett. {\bf 95},
226801 (2005)

\bibitem{haldane1} F. D. M. Haldane, Phys.
Rev. Lett. {\bf 61}, 2015 (1988)

\bibitem{haldane2} L. Sheng, D. N. Sheng, C. S. Ting, and F. D. M. Haldane
Phys. Rev. Lett. {\bf 95}, 136602 (2005).

\bibitem{kane2} C. L. Kane and E. J. Mele, Phys. Rev. Lett. {\bf 95},
146802 (2005).

\bibitem{yako1} V. M. Yakovenko, Phys. Rev. Lett. {\bf 65}, 251 (1990)

\bibitem{ng1} K.K. Ng and M. Sigrist, EuroPhys. Lett. {\bf 49}, 473
(2000).

\bibitem{hu1} C.R. Hu, Phys. Rev. Lett. {\bf 72}, 1526 (1994)

\bibitem{ks2} H-J. Kown, V.M. Yakovenko, and K. Sengupta Synthetic Metals {\bf 133-134},
27 (2003)

\bibitem{volovik2} G.E. Volovik, JETP Lett. {\bf 55}, 368 (1992).

\bibitem{srrefs} M. A. Tanatar {\it et al.}, Phys. Rev. Lett {\bf
86}, 2649 (2001); K. Izawa {\it et al.} {\it ibid} {\bf 86}, 2653
(2001); F. Laube {\it et al.}, {\it ibid} {\bf 84}, 1595 (2000); Z.
Q. Mao {\it et al.} {\it ibid} {\bf 87}, 037003 (2001).

\bibitem{rahul1} R. Roy, cond-mat/0603271 (unpublished).

\bibitem{ks3} K. Sengupta and V. M. Yakovenko, Phys. Rev. B {\bf 62}
4586 (2000).

\bibitem{kee1} H.Y. Kee, Y.B. Kim and K. Maki, Phys. Rev. B {\bf 61},
3584 (2000).

\bibitem{brey1} L. Brey and H. A. Fertig, cond-mat/0602505
(unpublished); N. Sinitsyn {\it et al.}, cond-mat/0602598
(unpublished).

\bibitem{ashoori1} R.C. Ashoori {\it et al.}, Phys. Rev. B {\bf
45}, 3894 (1992); G. Ernst {\it et al.} Phys. Rev. Lett. {\bf 79},
3748 (1997).

\bibitem{abanin1} D.A. Abanin, P. Lee, and L.S. Levitov,
cond-mat/0602645 (unpublished); N. M. R. Peres, F. Guinea, and A. H.
Castro Neto, PRB {\bf 73}, 125411 (2006).


\end{thebibliography}
\end{document}